\documentclass[prb,twocolumn,superscriptaddress]{revtex4}
\usepackage[dvips]{graphicx}
\usepackage{latexsym}
\usepackage{amsmath}
\begin{document}

\title{Quantum Criticality and Deconfinement in Phase Transitions
  Between Valence Bond Solids}

\author{Ashvin Vishwanath} \affiliation{Department of Physics,
  Massachusetts Institute of Technology, Cambridge MA 02139}

\author{L. Balents} \affiliation{Department of Physics, University of
  California, Santa Barbara, CA 93106-4060.}

\author{T. Senthil} \affiliation{Department of Physics, Massachusetts
  Institute of Technology, Cambridge MA 02139}

\date{\today}

\begin{abstract}
  
  {We consider spin-half quantum antiferromagnets in two spatial
  dimensions in the quantum limit, where the spins are in a valence
  bond solid (VBS) phase. The transitions between two such VBS phases
  is studied. In some cases, an interesting second order transition
  controlled by a fixed line with varying critical exponents is found.
  A specific example is provided by an antiferromagnetically coupled
  bilayer system on the honeycomb lattice where a continuous quantum
  phase transition can generically exist between two VBS
  phases. Furthermore, these critical points are deconfined, in the
  sense that gapped spin-$1/2$ spinon excitations emerge right at the
  transition. The low energy physics of this critical point (upto
  marginally irrelevant interactions) contains just a free
  quadratically dispersing `photon'. The phase structure on one side
  of this continuous transition is very intricate consisting of a
  series of infinitely closely spaced further transitions in a
  `devil's staircase' form. Analogies with previous examples of
  deconfined quantum criticality are emphasized.  Closely related
  transitions in single layer systems are explored. These are second
  order only at some multicritical points. The solvable
  Rokshar-Kivelson point of quantum dimer models of single layer
  systems is found to correspond to a non-generic
  multicritical point}
\end{abstract}

\maketitle

\newcommand{\fig}[2]{\includegraphics[width=#1]{#2}}
\newcommand{\be}{\begin{equation}}
\newcommand{\ee}{\end{equation}}

\maketitle

\section{Introduction}


Although the theoretical study of quantum phase transitions \cite{Hertz} has
recieved much attention over the last several decades, it remains a
source of rich and unexpected physics to this day. For example, recent
work\cite{Senthiletal,Senthiletal1} on the quantum criticality of
certain quantum magnets (spin-$1/2$ quantum
antiferromagnets on the square lattice) has shown that a direct and generically
continuous transition between the Neel state and a valence bond solid
state is possible. Such a transition violates `Landau's rules' of
classical phase transitions which prohibit a direct continuous transition between phases that
posess such different symmetries. 
Perhaps more interestingly
the critical theory is unusual and naturally expressed in terms of new
emergent, `deconfined' degrees of freedom that carry fractional spin,
along with an emergent $U(1)$ gauge field. This is despite the absence
({\em i.e} confinement) of such fractional spin excitations or the
associated gauge field in either phase. Such critical points were
dubbed `deconfined' quantum critical points (QCPs).
In this paper we shall examine another example of a ``deconfined''
quantum critical point which occurs in a different context i.e. between two VBS phases.

The existence of such deconfined QCPs may be surprising since the
gauge theories arising in condensed matter systems (apart from 
physical electromagnetism) are generally {\sl compact}, i.e. are
defined in terms of a periodic gauge connection $e^{ia}$ rather than a
single valued non-compact vector potential $a$.  In D=2+1, such gauge
theories are well known to be ``always'' (but see below) confining.
In particular, the simplest and apparently generic compact $U(1)$
gauge theory hamiltonian is
\begin{equation}
  \label{eq:CEM}
  H = \sum_r \left\{ \frac{1}{8\pi \epsilon}|{\bf E}|^2 +
    \frac{K}{2}|{\boldsymbol \Delta}\times{\bf E}|^2  - \gamma
  \cos {\boldsymbol \Delta}\times {\bf a} \right\},
\end{equation}
with integer valued ``electric field'' ${\bf E}$ and $2\pi$-periodic
vector potential ${\bf a}$ (canonically conjugate to one another)
defined on the links of a two-dimensional lattice.  Here
${\boldsymbol\Delta}$ is the lattice gradient, and ${\boldsymbol
  \Delta}\times$ is the lattice curl, defined as the gauge flux
through a plaquette.  In the natural cases arising in condensed
matter contexts, one imposes the Gauss law constraint ${\boldsymbol
  \Delta}\cdot{\bf E}=0$ (the ``even'' gauge theory) or ${\boldsymbol
  \Delta}\cdot{\bf E}_r=\epsilon_r$ on bipartite lattices, with
$\epsilon_r=\pm 1$ taking opposite signs on the two sublattices (the
``odd'' gauge theory).  In both cases, regardless of the values of
$\gamma$ or the ``dielectric constant'' $0<\epsilon<\infty$, the gauge theory
is always confining.  The value of the nominally ``irrelevant'' (in the
renormalization group sense) coupling $K$ is then immaterial.

Remarkably, Ref.~\onlinecite{Senthiletal}\ concluded that a deconfined
phase {\sl is} possible if a gauge theory similar to above is coupled to
gapless matter fields, which arise naturally in quantum antiferromagnets
at particular QCPs.  In this paper we study a different class of
transitions which provide other instances of `deconfined' quantum
criticality in quantum magnets.  Although in contrast to
Ref.~\onlinecite{Senthiletal}, they are not examples of Landau forbidden
transitions, they nevertheless display several interesting similarities
to those priors.  We consider transitions between two valence bond solid
(VBS) states: quantum paramagnets with a gap to spin excitations. In
either phase these spin-carrying excitations are conventional (though
gapped) spin-$1$ magnons or their composites. We study direct second
order transitions between two such phases where `deconfinement' obtains
right at the critical point. We show that the critical mode can be
viewed as a gapless deconfined $U(1)$ gauge field with a quadratic
dispersion.  Moreover the magnon excitations, that are sharply defined
gapped excitations on either side of the transition, break up into two
weakly interacting (gapped) spin-$1/2$ spinons. These spinons are
minimally coupled to the critical deconfined $U(1)$ gauge field. This
results in a weak interaction between the spinons so that they are
essentially free.

In contrast to the examples of deconfined quantum criticality
of Ref.~\onlinecite{Senthiletal} which are strongly interacting
\cite{MV}, the critical fixed points discussed here have a simple free
field description.  In particular, they may be considered as a limit of
the compact gauge theory in Eq.~(\ref{eq:CEM}) where the dielectric
constant $\epsilon \rightarrow \infty$.  At this point, the energy cost
of an emergent electric flux to quadratic order in ${\bf E}$ is
proportional to $|{\boldsymbol\Delta}\times {\bf E}|^2$ rather than
$|{\bf E}|^2$ as usual.  Despite the absence of gapless matter fields,
this can provide a realization of a deconfined QCP, provided the system
has sufficient (physically achievable) symmetries, as shown below.  We
emphasize that the appropriate physical symmetries occur naturally, for
instance in an antiferromagnetically coupled spin-$1/2$ honeycomb
bilayer.  Moreover, the QCP in this case is ``generic'', i.e. only one
physical parameter needs to be tuned (corresponding to tuning
$1/\epsilon$ through zero) to reach it.  While particular ``quantum
dimer'' models (see below) have provided some concrete realizations of
compact $U(1)$ gauge theories, many other microscopic mechanisms leading
to Eq.~\ref{eq:CEM}\ exist, and it is fruitful to regard the gauge
theory as the most general progenitor of such deconfined QCPs.

Closely related phenomena have been mentioned previously in the
literature\cite{Henley, MoSoFr} in the context of work on the quantum
dimer model\cite{qdm}.  The dimer model provides a caricature of
spin-gapped quantum paramagnetic phases and admits on certain lattices
a solvable point known as the Rokhsar-Kivelson (RK) point. On a square
or honeycomb lattice the RK point is known to be critical and to
separate two phases with very different VBS order.  An important
development was the field theoretic description of this RK point, that
was conjectured by Henley \cite{Henley}, and further elaborated in
\cite{MoSoFr}.  In this paper we will analyze the generic behaviour
expected in such a transition between VBS phases, and show that it
differs in important ways from the physics of the RK point.
In particular for single layer quantum spin systems on the square
lattice we argue that generically there is no second order transition
between two VBS phases.  It has been argued in Ref. \onlinecite{huse}
that a similar situation also obtains for the single-layer honeycomb
lattice, namely there is generically no second order transition. In
contrast as we show below such a second order transition with
`deconfined' criticality obtains on the bilayer honeycomb lattice with
antiferromagnetic interlayer coupling. In fact, a line of fixed points
with continuously varying exponents is obtained. The universal long
distance physics at one of these critical points corresponds to the
long distance physics at the RK point (upto logarithmic corrections
arising from a marginally irrelevant operator). Thus, although in
general there are significant differences from the generic transition,
the RK point in some cases can still provide useful information about the
universal critical properties of these generic transitions. A unique
feature of the RK point is that it occupies a very special place in
the phase diagram of the generic spin model, a point that will be
further discussed below.

We now summarize our results for the continuous quantum phase transition
between two VBS phases of spin-$1/2$ quantum antiferromagnets on the
{\em bilayer honeycomb} lattice. The precise nature of the neighbouring
phases themselves will be described shortly. There is a simple Gaussian
description of the critical theory which is parametrized by a
`stiffness' $K$ and has a dynamical scaling exponent $z=2$. In addition
there is a marginally irrelevant operator that will lead to logarithmic
corrections. Apart from the usual `thermal' operator needed to tune to
the critical point, the theory is found to have no other relevant
perturbations for a range of stiffnesses, leading to a fixed line with
continuously varying exponents depending upon $K$. Indeed, this is an
interesting example of a fixed line in $D=2+1$ dimensions, and is
closely related to three dimensional statistical physics models of the
Lifshitz point in certain liquid crystal systems \cite{Grinstein}.


Our results rely crucially on a `dual' description of the quantum
paramagnetic phases of antiferromagnets (with collinear spin
correlations) in terms of a sine-Gordon (or `height') field
$\chi$\cite{ReSaSuN,Frbook}.  It is constructed (see Section~\ref{sgsec})
by defining $E_i = \epsilon_{ij} \Delta_j \chi$, so that all physical
properties are invariant under the global shift $\chi\rightarrow
\chi+1$.  This formulation may be obtained in a number of different ways
which are briefly discussed in Section \ref{sgsec} below. In the
continuum limit appropriate near critical points between various
paramagnetic phases the Euclidean action of this model reads (for the
honeycomb lattice {\it bilayer} system):
\begin{eqnarray}
\label{sgcont}
S & = & S_0 + S_1 + S_{inst} \\
S_0 & = & \frac12 \int \! d^2x d\tau  \big\{ (\partial_\tau \chi)^2 + \rho (
{\bf \nabla}  \chi)^2
+K (\nabla^2 \chi)^2\big\}\\
S_1 & = & \int \! d^2x d\tau \, \frac u 4 |{\bf \Delta}\chi|^4 + \dots \\ 
S_{inst} & = & -\int \! d^2x d\tau \, \lambda \cos(2\pi \chi)
\end{eqnarray}
This continuum action is invariant under global integer shifts of the
$\chi$ field, as well as $\chi \rightarrow -\chi$ (${\bf E}\rightarrow
-{\bf E}$) which arises from symmetry under bilayer exchange. The
ellipses represent other terms (higher derivatives and higher powers)
which could be added that are consistent with the symmetries above. As
discussed below, the field $\chi$ may be interpreted as the dual of a
$U(1)$ gauge field. In this interpretation the $\lambda$ term describes
instanton events. The transition of interest occurs when $\rho$ changes
sign with $\rho = 0$ at the critical point.  Instantons are relevant in
either phase but will turn out to be irrelevant at the critical point
for some range of $K$.  In addition the quartic term $u (\vec \nabla
\chi)^4$ will be shown to be marginally irrelevant. 
Thus, the action
$S_0$ in Eqn. \ref{sgcont}  (which was recently studied along with
several interesting generalizations in Ref. \cite{Ardonne}) will
describe, upto marginally irrelevant terms, the fixed point action of
a generic quantum critical point between two VBS phases of the {\it
bilayer} honeycomb lattice antiferromagnet.  
When $\rho >0$ the
sine Gordon field (also known as the height field) $\chi$ has zero
`tilt' in the ground state.  It describes a featureless paramagnetic
state of the original quantum magnet, which may be caricatured as being
made of singlets formed on the interlayer rungs. However when $\rho < 0$
one expects a finite `tilt' i.e.  $\chi_r \sim {\bf Q \cdot r}$. Thus
the transition may be expected to occur between a VBS phase with zero
tilt of the height field and a VBS phase with a nonzero tilt of the
height field -- the tilt increasing `continuously' from zero on moving
away from the `deconfined' critical point.  Indeed while this general
expectation is more or less correct, the detailed picture is somewhat
more complicated owing to the relevance of the instanton operators away
from the transition. In fact, as described in section \ref{devil},
interesting structure obtains on the `tilted' side of the transition.
The background electric field (tilt) changes from its value of zero at
the critical point to a non-zero value some distance away, through a
fractal sequence of interleaving regions of constant tilt (confined
phases) and non-constant incommensurate regions (deconfined phases) of
zero width but finite total measure.  The resulting structure is known
as an `incomplete devil's staircase'.\cite{huse}

In contrast to the above picture, the RK point sits between a VBS phase
with zero tilt of the height field, and the staggered phase which has
the maximum tilt to the height field, i.e. with $2\pi{\bf Q}$ above one of
the six reciprocal lattice vectors of minimum length. These phases are
illustrated in Fig.~\ref{fig:honey}.  One can then ask whether the RK
point occupies a special place in the phase diagram of the generic
model. In fact, the well known property of the RK point that the ground
state in each tilt (winding number) sector are all degenerate indicates
that indeed it occupies a special place in the phase diagram. We have
already noted that the
RK point exhibits the same universal critical properties as the generic
model after tuning a few relevant and marginal operators. However, to
reach the special position occupied by the RK point in the phase
diagram, fine tuning of several dangerously irrelevant operators is
required. These operators do not affect critical properties but
determine the phase structure in the vicinity of the critical point.
The precise position of the RK point in the phase diagram is dicussed
below. Note, the generic phase structure can be recovered if we add to
the RK point an operator that (for example) corresponds to the quartic
term $S_1$ in equation (\ref{sgcont}). Moreover, this particular
operator will also generate logarithmic corrections to the RK point
correlation functions, which will then precisely match the one of the
generic critical points for the bilayer honeycomb system. Thus, generic
behaviour, both in terms of critical properties as well as phase
structure neighbouring the critical point can be obtained by adding this
one operator to the RK point Hamiltonian.



In the usual quantum dimer model, this fine tuning of the Hamiltonian
to access the RK point is achieved by keeping only terms that involve
a single plaquette. However, if we regard the dimer model as an
approximate description of some underlying quantum spin model, then in
general we need to include dimer kinetic and potential terms on
arbitrarily large-sized loops (but with coefficients that decrease
with increasing loop size). The exact degeneracy mentioned above then
obtains only if the dimer kinetic energy on every such loop is set
equal to the corresponding potential energy. Thus for an underlying
quantum spin model this exact degeneracy presumably requires
`infinite' fine-tuning.

A different way to draw a sharp distinction between the RK point and a
generic (two parameter tuned) multicritical point that allows for a
direct zero tilt to staggered (maximum tilt) state transition, is to
ask -- what phases can be accessed from these points by a small change
in bare parameters?  From the generic multicritical point (shown in
Figure \ref{RKpt}b) three states -- the staggered, zero tilt and an
infinitesimally tilted state -- can be accessed. From the RK point
however, in addition to the above three states, states with arbitrary
values of the tilt can be accessed (Figure \ref{RKpt}c). This is a
consequence of the exact degeneracy of states with different tilts at 
the RK point which gives rise to the special phase structure in its 
vicinity. 

The layout of this paper is as follows. In Section \ref{sgsec},
we discuss the sine Gordon representation of this problem from
different points of view, the non-linear sigma model approach of
Haldane, the easy plane deformation of the spin half magnet of
Lannert et al. \cite{Courtney}, and the familiar height
representation of the quantum dimer model. In Section
\ref{bilayerhoneycomb} we analyze the spin half quantum antiferromagnet
on the bilayer honeycomb lattice using this representation, and find a
stable fixed line controlling the transitions. Various properties of
these critical points are discussed, and the `devil's staircase' phase
structure is obtained. In section \ref{square}, the single layer
square and honeycomb lattice spin-half antiferromagnet is discussed,
where the generic transition between VBS phases is first order. In
section \ref{RK}, we discuss the well known RK points, and how they
fit into the general structure described in this paper.

\section{Sine Gordon representation for  paramagnetic phases of collinear quantum  antiferromagnets}
\label{sgsec}

Our analysis relies crucially on a formulation\cite{ReSaSuN,Frbook} of
the physics of quantum paramagnetic phases in terms of a sine Gordon
field theory on the dual lattice.  In this Section, we describe this
sine Gordon description and discuss its origin and interpretation from
several different perspectives which together provide considerable
insight. We will first discuss the single layer case and then move onto the double layer. 
The lattice Euclidean action for the sine Gordon model appropriate to a single layer has
the following structure:
\begin{eqnarray}
\label{sg}
S & = & S_1 + S_2 + ..... \\
S_1 & = &  \frac12 \int_\tau \sum_r (\partial_\tau \chi_r)^2 + \rho ( {\bf \Delta} \chi)^2
+K (\Delta^2 \chi)^2 \\
S_2 & = & -\int d\tau \sum_r \sum_{n = 0}^{\infty} \lambda_n \cos\left(2\pi n (\chi_r - \alpha_r)\right)
\end{eqnarray}
Here $r$ runs over the sites of the dual lattice. The symbol $ {\bf
\Delta}$ refers to a lattice derivative. The ellipses represent other
terms that are consistent with all the symmetries (both internal and
lattice) that could be added. In the second term $n$ is an integer
that runs from $0$ to $\infty$. The $\alpha_r$ are independent of time
but vary in a definite manner on spatial lattice sites. On the square
lattice, $\alpha_r = 0, 1/4, 1/2, 3/4$ on four sublattices. Thus
$\exp{(2i\pi \alpha_r)}$ oscillates rapidly on four sublattices. For a
spin model on the honeycomb lattice, the corresponding sine Gordon
theory is defined on the dual triangular lattice. In this case
$\alpha_r = 0, 1/3, 2/3$ on the three sublattices of the triangular
lattice so that $\exp{(2i\pi \alpha_r)}$ oscillates rapidly on three
sublattices.

Near the phase transitions of interest in this paper, the universal
physics is adequately described by a continuum limit of this
action. The oscillating phase factors due to the $\alpha_r$ means that
in the continuum the only terms that survive from $S_2$ are those with
$n = 0 (\mbox{mod}\, 4)$ for the square and $n = 0 (\mbox{mod}\, 3)$
for the honeycomb lattices respectively. 

The physical basis of the sine Gordon description may be understood in
many ways.  We will mainly discuss the square lattice -- the honeycomb
lattice is very similar.  

{\bf Gauge theoeretic/quantum dimer description:} First consider a description of
VBS phases in terms of quantum dimer models.  A dimer is to be
considered a caricature of a single valence bond and is taken to live
on the bonds of the original lattice.  For spin-$1/2$ systems, it is
natural to constrain the dimer Hilbert space by requiring that there
is exactly one dimer emanating out of each lattice site.  It has long
been appreciated that such quantum dimer models on bipartite lattices
can be fruitfully viewed as compact $U(1)$ gauge theories.  This is
understood as follows: First divide the bipartite lattice under
consideration into $A$ and $B$ sublattices. To each dimer we may
associate an integer-valued vector `electric field' ${\bf E}$ that
starts from an $A$ sublattice point and ends at a $B$ sublattice
point. In terms of these electric fields, the dimer constraint simply
becomes the Gauss law
\begin{equation}
{\bf \Delta}\cdot {\bf  E} = \pm 1
\end{equation}
where the $+$ sign is for the $A$ sublattice and the $-$ sign for the
$B$ sublattice.
Thus we see that the dimer Hilbert space is identical to that of a
particular compact $U(1)$ gauge theory with fixed background `charges'
$\pm 1$ on the two sublattices. Such theories were christened `odd
gauge theories' in Ref.~\onlinecite{MoSoFr}.

 Alternately such a gauge theoretic description of the disordered
 phases of the antiferromagnet can be directly obtained by starting
 with a slave particle ({\em eg.} Schwinger boson or fermion)
 description. Consider for instance a Schwinger boson representation.
 This has a $U(1)$ gauge redundancy associated with arbitrary phase
 rotations of the bosons at each site. In a spin gapped paramagnetic
 phase the Schwinger bosons can be integrated out and the physics is
 described by a compact $U(1)$ gauge theory. In this route too an odd
 gauge theory obtains.

There is a well-known duality mapping between compact $U(1)$ gauge
theories and sine-Gordon field theories -- see for instance Ref.~\onlinecite{Poly}.
The interpretation of the sine Gordon field is as follows. In the
absence of the compactness in the gauge theory, the total magnetic
flux is exactly conserved. There is a corresponding (topological)
global $U(1)$ symmetry. In the sine-Gordon representation this
becomes an ordinary global $U(1)$ symmetry that corresponds to an
arbitrary global shift of the sine-Gordon field. This symmetry is
present in the sine Gordon model if the coefficient $\lambda_n$ of
all the cosine terms is set to zero. Including compactness in the
original gauge theory allows for `instanton' events which destroy
flux conservation. Indeed the flux can change in multiples of
$2\pi$. These are precisely captured by including terms like
$\cos 2\pi n(\chi_r - \alpha_r)$ in the sine Gordon description.
Thus $e^{2\pi i \chi}$ corresponds to an instanton event at which
$2\pi$ units of gauge flux is created. The shift $\alpha_r$ is due
to the `oddness' of the gauge theory. Specifically the presence of
background charges in the gauge theory leads to Aharanov-Bohm
phases for the gauge flux which are encapsulated in the shifts
$\alpha_r$ -- for details see Ref \cite{subirrev}.

{\bf Semiclassical Description:} Much further insight is obtained by
a semiclassical perspective that is particularly appropriate if there
is considerable short-ranged Neel order.  Deep in the Neel phase the
long distance low energy fluctuations of the Neel vector are described
by the familiar quantum $O(3)$ non-linear sigma model field theory. To
correctly describe quantum paramagnetic phases it has been recognized
for some time now that this continuum field theory must be augmented
by appropriate Berry phase terms that are sensitive to the microscopic
spin at each lattice site and to the details of the lattice structure.
The Berry phase terms vanish for all smooth configurations of the Neel
field.  In two spatial dimensions note that such smooth configurations
allow for topologically non-trivial configurations known as skyrmions.
However as shown by Haldane the Berry phases are non-zero in the
presence of singular configurations -known as hedgehogs or monopoles -
in space-time. At the location of the monopole the skyrmion number
associated with the Neel field configuration changes.  That such
skyrmion tunneling events are allowed is a consequence of the presence
of a lattice in the microscopic spin model. The calculation of
Ref.~\onlinecite{hald88} shows that the Berry phases associated with a
single monopole (defined on the plaquettes of the original lattice)
oscillates from one plaquette to another with amplitude
$\exp{2i\pi\alpha_r}$ where $\alpha_r$ is as defined above on the dual
lattice sites (or equivalently on the plaquettes of the original
lattice).

The quantum paramagnetic state is associated with a proliferation of
such monopole events.  The Berry phases associated with a single
monopole event leads to it transforming non-trivially under lattice
symmetry operations.  Thus proliferation of single-strength monopoles
leads to broken lattice symmetry in the paramagnetic state.

To describe the different possible paramagnetic phases, it is
convenient to imagine integrating out all gapped spin-carrying
excitations and focus on the physics of the skyrmion fields and the
associated monopole events.  First consider a limit in which the
monopoles are ignored ({\em i.e} imagine tuning the monopole fugacity
to zero).  In this limit the skyrmion number is exactly conserved.
This corresponds to a hidden (topological) global $U(1)$ symmetry in
the absence of monopoles. The paramagnetic phase may then be thought
of as a condensate of these skyrmions so that this global $U(1)$
symmetry is spontaneously broken. The low energy excitations are
fluctuations of the phase of the skyrmion field and will be
gapless. Indeed the corresponding action may be identified with $S_1$
above with $\chi/2\pi$ being the phase of the skyrmion
field. Including monopole events leads to explicit breaking of this
global $U(1)$ symmetry.  Clearly in this picture the $S_2$ term
corresponds precisely to skyrmion creation events (which are the
monopoles) with the appropriate Haldane oscillating phases
encapsulated in the shift fields $\alpha_r$.

{\bf Easy Plane Limit:} Finally we briefly mention a derivation of the
action above in the easy plane limit of the original spin model. As
argued in Ref.
\cite{Courtney,sp} (see Ref, \cite{Senthiletal} for a physical
discussion), a very useful continuum description of easy plane
spin-$1/2$ magnets is provided by focusing on vortex fields (known
as `merons') in the (XY-like) order parameter. The dual action
takes the form
\begin{eqnarray}
\label{crtny}
{\cal L}& = & {\cal L}_{\psi}+ {\cal L}_v + {\cal L}_A + {\cal L}_{inst} \\
{\cal L}_{\psi} & = &
\sum_{a = 1,2}|\left(\partial_{\mu} - iA_{\mu}\right)\psi_{a}|^2 + r |\psi|^2 + u\left(|\psi|^2 \right)^2 \\
{\cal L}_v & = &
 v |\psi_1|^2 |\psi_2|^2 \\
{\cal L}_A & = & \kappa\left(\epsilon_{\mu\nu\kappa}\partial_{\nu}A_{\kappa}\right)^2 \\
{\cal L}_{inst} & = & - \lambda_4 \left(\left(\psi^*_1 \psi_2 \right)^4 + c.c \right).
\end{eqnarray}
Here $\psi_{1,2}$ represent the two meron fields that are minimally
coupled ( as is usual) to a non-compact $U(1)$ gauge field $\vec A$,
and $|\psi|^2 \equiv |\psi_1|^2+|\psi_2|^2$.
As described in Ref.~\onlinecite{Senthiletal} for the square lattice the last
term physically describes the monopole or instanton events discussed
above for the isotropic models. This continuum model has a global
$Z_4$ symmetry associated with $\pm (\pi/4 + m\pi/2)$ shifts of the
phase of $\psi_1$ and $\psi_2$ respectively ($m = 0,1,2,3$). In this
description valence bond solid phases correspond to $\langle \psi_1\rangle =
\langle\psi_2 \rangle  \neq 0$. In such a condensate, the gauge field $\vec A$
acquires a mass by the usual Anderson-Higgs mechanism and may be
ignored at low energies.  Furthermore the global $Z_4$ symmetry is
also broken -- thus a low energy description is provided by focusing on
the relative phase $\theta$ between $\psi_1$ and $\psi_2$.  Clearly
the theory has the same structure as the continuum limit of
Eqn. \ref{sg} and we identify $\chi= 2\pi \theta$. This discussion
readily generalizes to the honeycomb lattice -- the main difference is
that the monopole events are tripled (leading to $Z_3$ symmetry).

>From any one of these perspectives it is clear that (anti)vortices in
the $\chi$ field correspond to spin-$1/2$ spinon configurations in the
original spin model. Specifically we define a (anti)vortex as a point
in space around which
\begin{equation}
\int d{\bf l}\cdot {\boldsymbol \nabla} \chi = \pm 1
\end{equation}
For instance in the dimer model these correspond precisely to
points where the dimer constraint is violated ({\em i.e} to
monomers). Equivalently we note that skyrmions and spinons see
each other\cite{note1} as sources of $2\pi$ flux -- so that a spinon
configuration corresponds to a vortex in the skyrmion phase.

{\bf Bilayer Systems:} The discussion above is readily adapted to bilayer
systems. Specifically consider a bilayer spin-$1/2$ quantum
antiferromagnet on a square or honeycomb lattice. The symmetries of
the microscopic Hamiltonian now include the Ising-like layer exchange
symmetry in addition to $SU(2)$ spin rotation, time reversal and all
the lattice symmetries. This layer exchange symmetry will play an
important role in our analysis. Consider first the limit in which the
interlayer antiferromagnetic exchange on each rung is the largest
coupling. In this limit it is appropriate to first diagonalize the
`rung' Hamiltonian. For each rung, the ground state is a singlet and
the first excited state is a triplet. A useful model of such a bilayer
is to replace each rung by an $O(3)$ quantum rotor with the
Hamiltonian
\be
H = \frac{g}{2}\sum_i \vec L_i^2 - J \sum_{<ij>}\hat n_i \cdot \hat n_{j} + .....
\ee
Here $\hat n_i$ is a unit three component vector defined on each rung
(labelled by i) and $\vec L_i$ is the corresponding angular
momentum. The ellipses represent other short-ranged terms consistent
with the symmetries. The rotor vector $\hat n_i$ and the angular
momentum $\vec L_i$ have the same symmetry properties as the
difference and sum of the two microscopic spins on the rung at $i$
respectively. Under layer exchange we then have
\begin{eqnarray}
\hat n_i & \rightarrow & - \hat n_i \\
\vec L_i & \rightarrow & \vec L_i
\end{eqnarray}
Thus layer exchange symmetry implies that the rotor Hamiltonian be
invariant under the full group $O(3)$ of rotations (which includes
improper rotations). In addition time reversal is a separate symmetry
that is implemented by an antinunitary operator that changes the sign
of both $\hat n$ and $\vec L$. We remark that this must be contrasted
with single layer Heisenberg spin magnets. These can also be modelled
as quantum rotors but with appropriate monopoles placed at the origin
of the $\hat n$-sphere at each site. Now the presence of the monopoles
implies that the improper rotations of the rotor vector are no longer
symmetries. Thus these must be regarded as $SO(3)$ rotors. This
distinction will also be important for us below.

Consider now paramagnetic phases of the bilayer model. We will
specifically be interested in phases that obtain close to the strong
interlayer exchange limit where the $O(3)$ rotor description becomes
appropriate. As with the single layer systems discussed above it will
be convenient to obtain a gauge theoretic description of these
paramagnetic phases. This may be obtained by passing to a $CP^1$
description of the rotors in terms of spinon variables $z$. The $z$
fields are minimally coupled to a compact $U(1)$ gauge field $\vec a$
but unlike the single layer case the mean spinon number is zero per
site.  In a mean field description of paramagnetic phases the spinon
fields will be gapped. Beyond mean field, integrating out the gapped
spinons leads to a compact $U(1)$ gauge theory. The ultimate fate of
the spinons is determined by whether or not this gauge theory is
confined. Again in contrast to the single layer case the Gauss law
constraint of this gauge theory is simply
\be
{\bf \Delta}\cdot {\bf  E} = 0
\ee
with no background charges. Here $E_{ij}$ is the `electric' field
defined on the links of the honeycomb or square lattice. As usual this
is conjugate to the gauge field $a_{ij}$:
\be
[a_{ij}, E_{ij}] = i.
\ee
 What are the symmetries of this gauge theory? Clearly all the
 symmetries of the square or honeycomb lattice that forms each layer
 are also symmetries of the gauge theory. In addition the symmetry of
 the rotors under improper rotations (the layer exchange symmetry)
 implies that the gauge theory must be invariant under the discrete
 symmetry
\begin{eqnarray}
\label{ph}
E_{ij} & \rightarrow & -E_{ij} \\
a_{ij} & \rightarrow & -a_{ij}
\end{eqnarray}
This may be seen in several ways. For instance we note that the
magnetic field corresponding to the gauge field $a$ is precisely the
skyrmion density associated with the configuration of the $\hat n$
fields. The latter is odd under improper rotations of $\hat n$ (for
instance there is a well-known expression for the skyrmion density as
a trilinear in $\hat n$). Similarly the electric fields correspond to
the skyrmion currents which are likewise odd under improper
rotations. Parenthetically we note that under (the antiunitary) time
reversal the electric field is even while the gauge field is odd.

This compact $U(1)$ gauge theory is readily dualized to obtain a dual
sine-Gordon description in terms of the $\chi$ field. The dual action
takes the form
\begin{eqnarray}
\label{sgbl}
S & = & S_1 + S_2 + ..... \\
S_1 & = &  \frac12 \int_\tau \sum_r (\partial_\tau \chi_r)^2 + \rho ( {\bf \Delta} \chi)^2
+K (\Delta^2 \chi)^2 \\
S_2 & = & -\int d\tau \sum_r \sum_{n = 1}^{\infty} \lambda_n
\cos\left(2\pi n \chi_r \right)
\label{sgbl1}
\end{eqnarray} 
There are two important differences with the single layer case. First
the absence of background charges in the gauge theory implies that
there are no offsets $\alpha_r$ for the height fields $\chi_r$. Second
the discrete layer exchange symmetry implies that the action must be
invariant under $\chi_r \rightarrow - \chi_r$.

It is interesting to contrast the bilayer with a spin-$3/2$
antiferromagnet in a single honeycomb layer. In the latter the gauge
theory appropriate to the paramagnetic phase may be viewed as an
`even' gauge theory, {\em i.e} one where there are no background
charges in the Gauss law constraint. But nevertheless as the
microscopic model is not invariant under improper rotations of the
spin (or equivalently the rotor vector in a rotor description) the
gauge theory does not have the discrete symmetry of Eqn. \ref{ph}
associated with changing the signs of both $E$ and $a$.

\section{The Bilayer Honeycomb Lattice}
\label{bilayerhoneycomb}

In this Section, we specialize to the {\it bilayer}
honeycomb lattice, assuming the presence of strong interlayer
antiferromagnetic coupling. We will consider the lattice valence bond
solid phases and phase transitions of this spin half quantum
antiferromagnet using the sine Gordon description Eqn. \ref{sgbl}. It
will be convenient to explicitly write out the lowest order non-linear
terms that are allowed by the symmetries.  These take the form
\begin{equation}
\label{Sinteraction}
S_{int} = \int d\tau \frac u 4 |{\bf \Delta}\chi|^4 + \frac{v}{6}{\rm
  Re}[(\Delta_x+i\Delta_y)\chi]^6
\end{equation}
so that the full action is
\begin{equation}
\label{haction}
S = S_1 + S_2 + S_{int}
\end{equation}
with $S_{1,2}$ given in Eqn. \ref{sg}.  We have included in $S_{int}$
the $v$ term which gives the lowest order effect of the sixfold broken
rotational symmetry of the hexagonal lattice.  Note that it appears
only at {\sl sixth} order in $\Delta\chi$, and hence is nominally
irrelevant at the critical point.  It is nevertheless important for
$\rho<0$ (see below). Note that the symmetries of the bilayer system
include $\chi_r \rightarrow -\chi_r$ at each site $r$ which
corresponds to layer exchange. This symmetry forbids the appearance of
terms that are odd in $\chi$. Such terms are allowed in single layer
systems that do not have this symmetry - indeed Ref.~\onlinecite{huse}
identifies a cubic term that drives the transition first order.

\begin{figure}
\includegraphics[width=8.7cm]{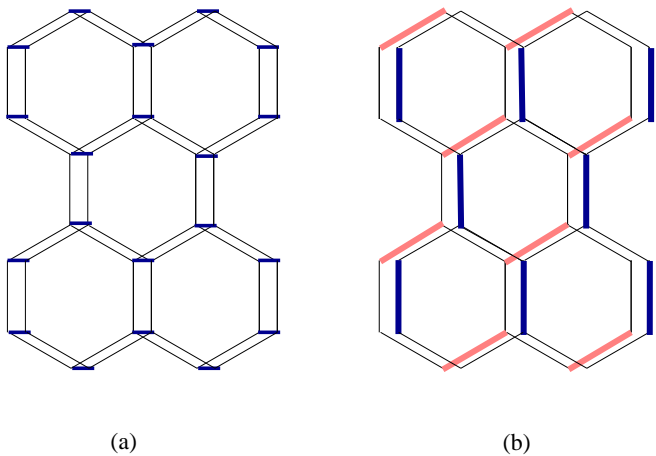}
\vspace{0.1in}
\caption{\label{fig:honey}
  Caricature of VBS phases on the bilayer honeycomb lattice. (a) The
  zero tilt state, with singlet bonds (thick lines) on the interlayer
  rungs. Note, this state does not break any lattice symmetry. (b) One
  of six possible maximally tilted (staggered) phases.}
\end{figure}
Let us for a moment consider how the various VBS phases arise from the
action (\ref{haction}). For $\rho>0$, the system would like to have zero
tilt ${\bf \Delta}\chi$ on the average. Moreover, we can ask what the
effect of the monopole tunneling term is, the first such term is the one
that inserts a single monopoles ($n=1$) of Eqn.\ref{sgbl1}, and it is
easily seen that this operator has long ranged correlations if $\rho \ne
0$, and is hence a relevant perturbation. The resulting phase will be
one where the height field is pinned at a uniform value, and this state
may be caricatured as one where the spins form singlets with their
partners in the other layer as shown in Fig. (\ref{fig:honey}a).

For negative values of $\rho$, a state with a finite tilt is expected,
i.e.
\begin{equation}
  \label{eq:tiltmf}
  {\bf k} = \langle 2\pi {\bf \Delta}\chi \rangle \neq 0.
\end{equation}
Ignoring the monopole operators for the moment, the system will choose a
tilt ${\bf k}_0$ whose magnitude for small negative $\rho$ is obtained
at the mean-field level by minimizing together the $\rho$ and $u$ terms,
i.e.
\begin{equation}
  \label{eq:tiltmfa}
  |{\bf k}_0| = 2\pi\sqrt{\frac{-\rho}{u}},
\end{equation}
with corrections of $O[v/u (-\rho/u)^{3/2}]$ from $v$ and other higher
order terms.  Fluctuation effects due to the marginality of $u$ will
slightly enhance $|{\bf k}_0|$ by a logarithmic factor of little
importance.  The direction of the vector ${\bf k}_0$ is, however,
determined by the sign of $v$.  In particular, the six discrete
directions with $k_{0x}+ik_{0y}=|{\bf k}_0| e^{i2\pi m/6}$ or
$k_{0x}+ik_{0y}=|{\bf k}_0| e^{i2\pi (m+1/2)/6}$, with $m=0\ldots 5$ are
preferred for $v<0$, $v>0$, respectively.  The added effects of
monopoles, which will modify the true tilt vector to ${\bf k} \neq {\bf
  k}_0$, however, will have to be carefully considered discussed in the
last part of this section.

In any case, a phase transition between a VBS with zero average tilt,
and one with nonzero average tilt, corresponds to taking $\rho$ from
positive to negative values. The phase transition that lies between
requires that we look at the theory with $\rho=0$.

\subsection{Stability of the Fixed Line Controlling the Transition}

The critical theory for such a VBS transition is then proposed to be:
\be
\label{hcritical}
S_c =\int d\tau \sum_r\frac12\{(\partial_\tau \chi)^2 +{K}(\Delta^2 \chi)^2 \}
\ee
we need to check that this simple Gaussian theory is stable against
switching on a small monopole tunneling (term $S_2$ in the action
(\ref{sgbl})) and quartic interactions ($S_{int}$ of equation
(\ref{Sinteraction})).  Again, we consider correlators of the single  monopole tunneling event, the $n=1$ term. This now has
the following power law decay in space:
\be
\langle e^{i2\pi\chi_r(0)}e^{-i2\pi\chi_0(0)}\rangle \sim \frac1{r^{\frac{\pi}{\sqrt{K}}}}
\ee
which implies that the monopole insertion operator is irrelevant
if $\frac{\pi}{2\sqrt{K}}>4$, or equivalently
$0<K<(\frac{\pi}8)^2$. Thus, there is a line of fixed points, with
different exponents, that are parametrized by $K$ and are all stable
against switching on weak monopole tunneling. Indeed this is very
similar to the line of fixed points obtained in $D=1+1$ in a variety
of systems such as the spin half XXZ chain. Note however that our
theory has dynamical exponent $z=2$.

We now consider the effect of the quartic interaction term, $S_{int}$
in Eqn. \ref{Sinteraction},
which by naive power counting is marginal at these fixed points. Since
we will be looking at values of $K$ for which the monopole tunneling
events are irrelevant, they are disregarded in the discussion below.
We consider a continuum model of the critical theory
(\ref{hcritical}), with modes restricted to wavevectors below a certain
cutoff $\Lambda$,We perform a one loop RG, assuming that we start with
a small value of the interaction parameter $u$, and study its flow on
integrating out the large wave-vector modes
$\Lambda(1-dl)<|k|<\Lambda$. After the appropriate rescaling to keep
the quadratic term (\ref{hcritical}) invariant, we obtain the
following flow equation for the quartic coupling\cite{Grinstein}: 
\be
\frac{du}{dl} = -\frac9{16\pi K^\frac32}u^2.
\label{hRG}
\ee
This implies that a quartic coupling with $u>0$ is marginally
irrelevant, and the coupling flows back to zero logarithmically with
distance. Therefore the long distance physics in this case will be
controlled by the critical action (\ref{hcritical}), with logarithmic
corrections arising from this marginally irrelevant operator. Thus the
critical points are stable towards turning on a quartic interaction
for $u>0$. (For $u<0$ however, the quartic coupling is relevant, and
the transition is very likely driven first order -- as is indeed already
the case in mean field theory for negative $u$).

\subsection{Properties of the Transition}

The analysis above has established the presence of a fixed line
controlling the transition between VBS phases with zero tilt and those
with a non-zero tilt of the height field. In this subsection, we
discuss some properties of this fixed line. We first note that the
irrelevance of the monopole tunneling terms implies that the global
symmetry of the continuum theory is enlarged to $U(1)$. This
corresponds to invariance under arbitrary global shifts of the height
field $\chi$. In terms of the $U(1)$ gauge theory (whose dual is the
sine Gordon theory) this implies that the compactness is
asymptotically irrelevant all along this fixed line. Indeed as shown
in Appendix \ref{pht} the free Gaussian action is readily seen to
describe a quadratically dispersing gapless photon in the gauge theory
representation. The gauge flux of the theory is then conserved. This
signals deconfinement. However the monopoles are important to
correctly describe the physics of the phases on either side of the
transition -- thus they represent `dangerously irrelevant'
perturbations.

If we ignore the marginally irrelevant quartic term the fixed point
theory has a free field form. Thus all aspects of the critical
behavior while non-trivial are eminently tractable. For instance it
should be possible to compute real time dynamical correlators of, say,
the operator $e^{i2\pi  \chi_r(t)}$ at non-zero temperatures in the scaling
limit. We will however not pursue this here.

It is instructive to ask about the fate of the gapped spin-carrying
excitations right at this critical fixed line. As mentioned in Section
\ref{sgsec}, the presence of a spin-$1/2$ spinon at some spatial site
leads to a vortex in the height field $\chi$. Away from the critical
point, in (for instance) the zero tilt VBS, the relevance of the
monople tunneling terms lead to pinning of the height field 
which implies that there is a huge energy cost that
increases linearly with system size for these vortices. More precisely
consider a pair of spinons (on opposite sublattices of the original
lattice). This generates a vortex-antivortex pair in $\chi$. The
energy cost for separating this pair by a distance $R$ grows linearly
with $R$ away from the critical point. Thus we have (linear)
confinement of spinons and the elementary spin-carrying excitations
have spin-$1$. But right at the critical fixed line the enlargement of
the symmetry to $U(1)$ implies that vortices in the $\chi$ field
are cheap. Ignoring the quartic perturbation, an elementary
computation shows that the energy cost of a vortex is finite
independent of system size. Including the quartic term leads to a weak
$1/R^2$ interaction (up to logarithmic corrections due to the marginal
irrelevance of $u$) between a vortex-antivortex pair
separated by distance $R$ -- this then is the interaction between two
spinons (on two opposite sublattices) separated by a distance
$R$. Thus as expected the spinons are deconfined and free to propagate
above a spin gap.

Within the approximation of ignoring the weak interaction between the
two spinons, the magnon spectral function $A(\vec k, \omega)$ at the
gap edge is readily calculated.  One finds a sharp step $A(\vec k,
\omega) \sim \theta(\omega - \Delta)$ where $\Delta$ is the spin
gap. Thus the magnon spectral function has no quasiparticle peak and
is anomalously broad.

\subsection{Devil's Staircase}
\label{devil}

We now consider the behavior on the ``tilted'' side of the Lifshitz
point, in which the compact gauge theory is expected to have a
non-vanishing background electric field.  The neighborhood of the
Lifshitz point has been argued in Ref~.\onlinecite{huse}\ to realize
an ``incomplete Devil's staircase''.  While the essential features
have already been sketched in Ref.~\onlinecite{huse}, we
recapitulate them here for completeness and to present a few
additional points not mentioned therein.  In particular, we will
describe the thermal transitions of the commensurate tilted VBS states
within the devil's staircase, and point out two distinct types of low
energy excitations within these phases.

Let us first think generally about the nature of phases with some
finite and non-zero background electric field, i.e. not in the direct
vicinity of the putative critical point.  At this stage we do
not take any continuum limit, working with a field $\chi_r$ defined on
the discrete lattice sites (of the dual square or hexagonal lattices).
If one neglects at first the compactness of the gauge theory, i.e. the
terms breaking continuous translational symmetry of $\chi_r$, then one
may write $\chi_r = {\bf k}_0\cdot{\bf r}/2\pi + \delta \chi_r$, with
$\delta\chi_r$ describing fluctuations around the putative average
tilt ${\bf k}_0$ determined as in Sec.~\ref{bilayerhoneycomb}\ by minimizing
with respect to ${\bf k}_0$ the non-monopole terms in the action.  One
should regard ${\bf k}_0$ as the continuously varying ``tilt'' the
system would have were there no cosine (monopole) terms.  We will see
that the true tilt, i.e.  $\langle {\bf\Delta} \chi\rangle={\bf k}$,
is close but not generally equal to ${\bf k}_0$.  The fluctuations of
$\delta\chi_r$ will then be described by a theory of the form
\begin{eqnarray}
  \label{eq:tilt_flucts}
  S_{tilt}  & = & \int\! d\tau\, \sum_r \, \frac12 \{ |\partial_\tau
  \delta\chi_r|^2 + \tilde\rho |{\bf \Delta} \delta\chi_r|^2 \}
  \nonumber \\ && - \sum_n \tilde\lambda_n
  \cos[2\pi n\delta\chi_r + n{\bf k}_0\cdot{\bf r}],
\end{eqnarray}
Note that, unlike at the critical point, the fluctuations of
$\delta\chi_r$ at the quadratic level have a non-vanishing
(renormalized) stiffness $\tilde\rho$.  Hence the fluctuations of
$\delta\chi_r$ will be bounded, and any non-oscillatory cosine term
breaking the continuous translational symmetry of $\delta\chi_r$ will
``pin'' it, however weak.  This pinning corresponds to confinement in
the original gauge theory, and a VBS phase in the dimer model.  We
note in passing that, actually, depending upon the value of ${\bf
  k}_0$ and anisotropies in the original action, $\tilde\rho$ can be
replaced by a more general tensor.  Again, this complication does not
modify any of the qualitative results of this section, and so will be
ignored.

For a ``generic'' value of ${\bf k}_0$ {\sl all} the cosines oscillate
since ${\bf k}_0\cdot{\bf r}$ will be an irrational multiple of
$2\pi$.  There are, however, an infinite {\sl dense} set of values of
${\bf k}$ for which ${\bf k}\cdot{\bf r}$ is a rational multiple of
$2\pi$ (for all ${\bf r}$).  In this case, there will be some minimal
value of $n$ for which the $\lambda_n$ term does not oscillate (for
$\delta\chi_r=0$).  Clearly, at these special values of ${\bf k}$,
this cosine term is relevant and the system is in some confined VBS
phase.  Furthermore, any {\sl irrational} ${\bf k}_0$ is arbitrarily
close to one of these rational ${\bf k}$ values, so that, although the
cosines in general oscillate on the lattice, some of them oscillate
extremely slowly.  Since in general the cosines are {\sl not}
infinitesimally weak (i.e. the $\lambda_n$ are finite and non-zero), a
sufficiently long wavelength oscillation of the cosine term could
potentially ``pin'' the $\delta\chi_r$ field even in such cases.  To
see whether this occurs, let us suppose the $n^{\rm th}$ cosine term
oscillates weakly, i.e. $e^{i n{\bf k}_0\cdot{\bf r}} =
e^{in\delta{\bf k}\cdot{\bf r}}$ for all lattice vectors ${\bf r}$,
with $n|\delta{\bf k}|\ll 2\pi$.  Then, keeping only this cosine term,
we have approximately
\begin{eqnarray}
  \label{eq:tiltflucts2}
 && S_{tilt} =  \int\! d\tau\, \sum_r \, \frac12 \{ |\partial_\tau
  \delta\chi_r|^2 + \tilde\rho |{\bf \Delta} \delta\chi_r|^2 \}
  \nonumber \\
&&
 - \!\tilde\lambda_n \!
  \cos[2\pi n\delta\chi_r \!+\! n\delta{\bf k}\cdot{\bf r}].
\end{eqnarray}
Since the fluctuations of $\delta\chi_r$ are bounded, we may estimate
the effects of the cosine term by ignoring these fluctuations and
minimizing the action.  The minimal action configurations are clearly
constant in imaginary time, $\partial_\tau \delta\chi_r=0$.  Roughly,
then, the field $\delta\chi_r$ can either be constant, minimizing the
stiffness term but gaining no (lowering of the) energy from the
cosine, {\sl or} it can choose to tilt slightly to take advantage of
the cosine term, costing some energy from the stiffness.
Dimensionally, the latter will be favorable if $\tilde\lambda_n
\gtrsim \tilde\rho |\delta{\bf k}|^2$.  Now $|\delta{\bf k}|$ can be
made arbitrarily small by increasing $n$ (typically decreasing as
$1/n$), so this inequality will be satisfied if the $\tilde\lambda_n$
do not decrease too rapidly (i.e. faster than $1/n^2$) with $n$.
However, it is perfectly conceivable that the $\tilde\lambda_n$ terms
{\sl do} decrease faster than $1/n^2$, and in this circumstance, there
will be incommensurate values of ${\bf k}_0$ for which the $\chi$
field remains unpinned, and {\sl all} monopole terms remain
irrelevant.  This has been argued to be the case in the immediate
neighborhood of RK points in Ref.~\onlinecite{huse}.  Note that even
so, there are commensurate pinned states {\sl arbitrarily} close to
this incommensurate state.  The self-similar succession of various
commensurate and incommensurate states is the incomplete Devil's
staircase mentioned above.  The term ``incomplete''
indicates that the incommensurate unpinned regions exist (and have
finite measure, as can also be argued).   

The same arguments apply to the neighborhood of the VBS transitions
discussed here, and we will sketch the reasoning in order to make a
few more observations.  Note that in these cases (e.g. for the
honeycomb bilayer), the Lifshitz point itself is generic, i.e. can be
potentially observed in a physical system by varying only one
parameter.  

First, we comment on a minor subtlety.  A na\"ive analysis of the
continuum field theory, Eq.\ref{sgcont}, would suggest that on the
tilted side of the Lifshitz point, the tilt increases {\sl smoothly}
from zero.  This appears to be the case since at the Lifshitz point,
all cosine (monopole) operators are irrelevant.  One the tilted side
of the transition, in fact the tilt does not increase smoothly, but in
the staircase fashion.  This occurs because irrelevant cosine
operators at the Lifshitz point become relevant on the tilted side of
the ``transition'', i.e.  these operators are {\sl dangerously
  irrelevant} in renormalization group parlance.  In fact, the na\"ive
smooth increase in slope (background electric field of the gauge
theory) occurring on this side of the transition is replaced by a
slope which contains piecewise constant and incommensurate regions,
which become more and more closely spaced as the Lifshitz point is
approached, forming an infinite sequence that approximates the na\"ive
continuous curve (of e.g. $|{\bf k}|$ versus $\rho$) arbitrary well if
one looks arbitrarily close to the Lifshitz point.  

A full description of the devil's staircase is beyond the scope of
this paper.  It is instructive and indicative of the general structure
to consider some simple ``families'' of plateaus in the tilt ${\bf k}$
that obtain near the critical point.  In general, the condition for a
plateau is that $\cos(n{\bf k}\cdot{\bf r})$ does not oscillate on the
dual hexagonal lattice.  This condition is equivalent to requiring
that $n{\bf k}$ is a reciprocal lattice vector of the hexagonal
lattice.  An arbitrary reciprocal lattice vector can be written as
$n_1{\bf b}_1+n_2{\bf b}_2$, with ${\bf b}_1=(2\pi,2\pi/\sqrt{3})$,
${\bf b}_2=(0,-4\pi/\sqrt{3})$.  Hence, the condition on ${\bf k}$ is
\begin{eqnarray}
  \label{eq:kcomm}
  \vec{k}& = &
  \left(\frac{n_1}{n}\right){\bf b}_1
  +\left(\frac{n_2}{n}\right){\bf b}_2 
\label{eq:kcomm2}
\end{eqnarray}
where $n,n_1,n_2$ are integers, and the pairs $n_1,n$ and $n_2,n$ can
be taken to be relatively prime.  The ``strongest'' such commensurate
tilts are those with minimal $n$, since these correspond to
$n$-monopole events, which become less relevant as $n$ increases.

In understanding the behavior for small ${\bf k}_0$, we need to
investigate those commensurate tilts for which ${\bf k}$ is small but
non-zero.  Clearly for Eq.~\ref{eq:kcomm2}\ this will occur for large
$n$.  Since the two vectors ${\bf b}_1,{\bf b}_2$ are linearly
independent, the coefficients of both must be small for ${\bf k}$ to
be small.  Since the two numerators in these coefficients are
integers, for a given small magnitude $|k|\ll 1$, one clearly then
needs at least $n \gtrsim |k|^{-1} \gg 1$. Larger values of $n$ can
also yield the same $|k|$, by increasing the numerators accordingly.
However, the largest plateaus in tilt (coming from the most relevant
cosines with minimal $n$ for a given $|k|$) will be those
corresponding to $n \sim |k|^{-1}$.

Systems with ${\bf k}_0$ sufficiently near each of these values will
be pinned and form a ``plateau'' in ${\bf k}$.  How wide is
this plateau?  Let us suppose the {\sl putative} tilt ${\bf k}_0
\approx {\bf k}$.  By scaling, the correlation length $\xi \sim
1/|k_0| \sim n \gg 1$
In this situation, we
must account for the renormalization of the relevant cosine term by
the fluctuations on scales less than $\xi$.  From standard
renormalization group methods, one expects the renormalized
coefficient
\begin{equation}
  \label{eq:lamren}
  \tilde\lambda_n \sim \lambda_n \xi^{-n^2 \Delta},
\end{equation}
where $\Delta=\pi/(2\sqrt{K})$ is the scaling dimension of the
one-monopole term.  Note that, from this reasoning, the
$\tilde\lambda_n$ decrease extremely rapidly with $n$, hence from the
above argument, incommensurate phases are possible.

In addition, for large $\xi$, the renormalized
stiffness is small, $\tilde\rho\sim \rho/\xi^2$.  On scales longer than
$\xi$, the $\chi$ field is essentially non-fluctuating, so further
renormalization can be neglected.  The criteria for the system to be
pinned at ${\bf k}$ is then  $\tilde\lambda_n \gtrsim \tilde\rho |\delta
k|^2$.  Hence the width of the plateau is
\begin{equation}
  \label{eq:deltak}
  |\delta k| \lesssim \sqrt{\frac{\lambda_n}{\rho}} \xi^{1-n^2
    \Delta/2}.
\end{equation}
Clearly these commensurate plateaus are very narrow near the Lifshitz
point.  Similar estimates were obtained in Ref.~\onlinecite{huse}.

Each of the commensurate VBS phases breaks the discrete translational
symmetry of the lattice, and thus must undergo a symmetry-restoring
transition as temperature is increased.  At non-zero temperature $T$,
these states will be truncated to those commensurate VBS
phases whose $T_c^{'s}$ are larger than $T$.  In particular, consider
a commensurately tilted VBS state driven by the $n$-monopole fugacity
in the ``center'' of its plateau, i.e. for ${\bf k}={\bf k}_0$.  In
this case, we may take $\delta{\bf k}=0$ in
Eq.~(\ref{eq:tiltflucts2}), and the system is described simply by a
commensurate sine-Gordon model.  At $T>0$, we may neglect all but the
zero Matsubara frequency mode and set $\partial_\tau\chi=0$,
encompassing the quantum effects by using the renormalized parameters
$\tilde\rho,\tilde\lambda_n$ above.  The symmetry-restoring transition
is simply the roughening transition of this classical two dimensional
sine-Gordon model, which has
\begin{equation}
  \label{eq:tcclass}
  k_B T_c = \frac{2\tilde\rho}{\pi n^2} \sim \frac{1}{n^2 \xi^2}.
\end{equation}
For the strongest plateaus, recall $\xi \sim 1/|k|$ and $n\sim 1/|k|$,
hence one has for these plateaus $k_B T_c \sim |k|^4$.  The other
plateaus have even smaller critical temperatures.  Since all the
critical temperatures vanish rapidly as $|k|\rightarrow 0$, the infinite
set of plateaus in the devil's staircase is replaced by a finite subset
at any non-zero temperature.

One may also estimate the excitation gap in the associated commensurate
VBS phase, simply by expanding the sine-Gordon term to produce a
$(\delta\chi)^2$ ``mass'' term.  This gives a gap $E_g^{VBS} \sim n
\sqrt{\tilde\lambda_n} \sim n\sqrt{\lambda_n} \xi^{-n^2\Delta/2}$.  It may be
somewhat surprising that a state with an exponentially small gap can
have such a relatively {\sl high} critical temperature (power law in
$1/n$ from Eq.~(\ref{eq:tcclass}) ).  The physics of this is that the
excitation with an exponentially small gap consists of only small
fluctuations of $\delta\chi$ (and hence the gauge electric field) which
do not perturb the long-range order of the VBS state.  Indeed, since the
VBS phases are states of {\sl discrete} broken symmetry, the excitations
which do disturb this order by connecting the different symmetry-related
ground states are ``droplet''-like.  In the sine-Gordon language, such a
droplet may be thought of as a domain wall in $\delta\chi$,
which is wrapped around to form a compact ``island'' inside which
$\delta\chi$ is shifted by $\pm 2\pi/n$.  In particular, the
proliferation of such thermally excited droplets ultimately will
``depin'' the $\delta\chi$ field and destroy the VBS order above some
temperature.  The minimal radius of such an island is the domain wall width
itself (since this is larger than the other natural cutoff, the
correlation length).  The energy for such a droplet is therefore given
by integrating the exponentially small sine-Gordon term term over an
exponentially large area of the size of the domain width (in reality there
is also a comparable contribution from the $\tilde\rho$ term).  These
two factors compensate to give the relatively large energy determining
$T_c$.  The upshot of this argument is that the states with
exponentially small excitation gap near the zero tilt QCP are {\sl not}
associated with the typical classical droplet excitations of a VBS
state, but rather are evidence of the gapless photon mode obtaining
precisely at the QCP.

A comment on the above discussion is in order.  Within the sine-Gordon
treatment, at {\sl any} temperature above the ``roughening''
temperature, $\delta\chi$ behaves as a free scalar field, and vertex
operators $\exp{2\pi i n \delta\chi}$ exhibit power law correlations.
Ultimately, this can be tracked down as an artifact of the pure gauge
theory.  In particular, any matter fields included in the model, even
gapped ones, correspond as discussed earlier to vortices in the $\chi$
(or $\delta\chi$) field.  At sufficiently high temperatures, these
vortices will certainly unbind.  However, for $n>4$, it is known that
such sine-Gordon theories exhibit a ``floating'' phase in which power
law correlations persist, above the roughening temperature $T_c$ and
below the Kosterlitz-Thouless temperature $T_{KT}$ above which
vortices unbind.  Here $k_B T_{KT}= \tilde\rho/(8\pi)$.  Note that one
has then
\begin{equation}
  \label{eq:tcrat}
\frac{T_{KT}}{T_c} = \left(\frac{n}{4}\right)^2, \qquad {\rm for}\, n>4.
\end{equation}
Hence the VBS phases very close to the Lifshitz point will have
long-range VBS order only at very low temperatures $T<T_c\sim |k|^4$ but
quasi-long-range VBS order up to much higher temperatures $T<T_{KT}\sim
|k|^2$.  Moreover, in the region with quasi-long-range order, there are no
plateaus in the tilt.

In gauge theory language the roughening transition at $T_c$ may be
associated with the thermally driven deconfinement transition of pure
gauge theories. Indeed, electric field correlators in this rough phase
(gradients of the height field) fall off as the inverse square of the
distance -- so this phase may be thought of as a `thermal Coulomb
phase'. The gauge charged spinons are logarithmically interacting in
this phase and are bound into gauge neutral pairs until they ionize at
$T_{KT}$ leading to a plasma of gauge charge that destroys the long
range electric field correlations of the thermal Coulomb (or rough)
phase. This transition is also studied in detail in \cite{MV}, in
theories with a noncompact gauge field and SU(2) symmetric spion
fields. While the thermal deconfinement transition of gauge theories
is generally expected only in the absence of matter with unit gauge
charge (spinons), in two spatial dimensions the logarithmic form of
the Coulomb interactions is strong enough to bind the spinons into
gauge neutral pairs and hence the transition survives the inclusion of
spinons.

\section{The single layer honeycomb lattice}
In contrast to the situation analysed above for the bilayer honeycomb aniferromagnet, in a single layer the appropriate 
lattice sine Gordon model has non-zero offsets $\alpha_r$ for the $\chi$ fields on the three sublattices of the dual triangular lattice. 
Thus as explained in Ref.~\onlinecite{huse} the $\chi_r \rightarrow -\chi_r$ transformation becomes a symmetry only when combined with inversion or a $\pi/3$ 
rotation. This then leads to the possibility of a cubic invariant in the sine Gordon action which drives the transition first order. 

It is interesting to ask about the situation with spin-$3/2$ antiferromagnets on a single layer honeycomb lattice. In this case there are no offsets for the 
$\chi_r$ in the sine Gordon description of the paramagnetic phase. Nevertheless $\chi_r \rightarrow -\chi_r$ symmetry (without inversion or $\pi/3$
rotation) is not expected as an exact symmetry of the action. This is not required by any of the microscopic symmetries of the underlying lattice spin model. 
Hence we expect that a cubic term will still be allowed in the field theory and a first order transition will result.

\section{The Square Lattice}
\label{square}

We now perform the same analysis for the transition in the case of the
spin one half quantum antiferromagnet on the square lattice. In
contrast to the situation on the honeycomb lattice, we will find no
generic continuous transition between VBS states (in both single and bilayer cases). 

As before we consider a generic continuum theory that is consistent
with all the lattice and internal symmetries. For the square lattice
this takes the form
\begin{eqnarray}
\label{scritical}
S & = & S_c + S_{int} +S_{mon} \\
S_c &=& \frac12\int (\partial_\tau \chi)^2 + K[|\nabla^2 \chi|^2 +
4\sigma (\partial_x^2 \chi)(\partial_y^2 \chi)]\\
S_{int} &=& \int \frac{u}4 [(\partial_x \chi)^4+ (\partial_y
\chi)^4]+\frac v2 (\partial_x \chi)^2(\partial_y \chi)^2
\label{squartic}
\end{eqnarray}
with $\sigma>-1$. The isotropic point corresponds to $\sigma=0$ and
$u=v$. The last term ($S_{mon}$) represents quadrupled monopole
events. Note in particular the presence of the couplings $\sigma,v$
which are allowed by the square lattice symmetry. Consider the
critical theory given by the quadratic piece of the action $S_c$ and
determine its stability against the inclusion to small monopole
tunneling. In this case we need to consider quadrupled monopoles, as
discussed in section \ref{sgsec}, and once again it is possible to
find a range of $K, \sigma$ for which monopole tunneling is
irrelevant.

Next, one must consider the stability of the Gaussian fixed point
described by $S_c$ to turning on the quartic interaction
(\ref{squartic}). Again, we perform a one loop renormalization group
analysis to determine the fate of these couplings. The result of
integrating the high wavevector modes $\Lambda(1-dl) <|k|<\Lambda $
and rescaling, are the following RG equations:
\begin{eqnarray}
\label{RG1}
\frac{dU}{dl} &=& -3[9\alpha U^2+2UV +\alpha V^2], \\
\frac{dV}{dl} &=& -9[U^2+2 \alpha UV + V^2],
\label{RG2}
\end{eqnarray}
where we have used the scaled variables $U=u {\mathcal A}^{-1}$, $V=v
{\mathcal A}^{-1}$, and the scale factor ${\mathcal
A}=(1+\sigma+\sqrt{1+\sigma})32\pi K^{\frac32}$. At the isotropic
point, these equations are identical to (\ref{hRG}), and preserve
$u=v$. The anisotropy of the quadratic part of the action is present
in the parameter $\alpha$, which is unity at the isotropic point but
otherwise is given by:
\be
\alpha = \frac13[\frac{2\sigma+1-\sqrt{1+\sigma}}{\sqrt{1+\sigma}-1}]
\ee
thus $\alpha \in [1/3,\infty)$.

We now analyze the RG equations (\ref{RG1},\ref{RG2}), and show that
they imply that the critical Gaussian theory $S_c$ is generally
unstable in the presence of the two quartic operators. If these were
stable critical points, then there should be a region in the $u,\,v$
plane where the flows end up at the origin. However we will show that
there is only a single line in the entire $u,\,v$ plane, where the
couplings flow into the origin. This implies that stability is only
attained on a set of measure zero points. For a general choice of the
quartic coupling, the flows run away to large negative values of $u$
suggesting a first order transition. In order to show this property of
the RG equations (\ref{RG1},\ref{RG2}), we construct a function
$E(U,V)$ that is invariant along any RG trajectory
i.e. $\frac{dE}{dl}=0$. Contour lines of this function then represent
the RG flows, and we will see that there is only a single contour
connected to the origin. This function is most conveniently
represented in terms of the rotated coordinates, $U_+=\alpha U + V$,
$U_-=U - \alpha V$. Then,
\be
E=\frac{U_-^3}{(\alpha^2+3)U_-^2 + 4\alpha U_+ U_- + (3\alpha^2+1)U_+^2},
\ee
which can be checked to be invariant under the RG flows. It may be
seen that the origin must correspond to $E=0$, by approaching it in
any direction. The only other points for which the invariant function
vanishes is the line $U_-=0$, which corresponds to $u=\alpha v$. These
are the only points that could flow to the origin. Apart from
this set of measure zero, none of the other points in the $u, \, v$
plane ever reach the origin under the RG flows but rather flow to
regions with large negative values of $u$.  Thus, for the case with
square symmetry, the Gaussian critical theory is generically unstable
- the flows suggest a first order transition in the absence of special
fine-tuning.

\section{RK Points}
\label{RK}

The quantum dimer model Hamiltonians studied in \cite{qdm} were shown
to have a special point - the RK point - at which the wavefunction is
an equal superposition of all dimer configurations. Equal time
correlation functions can then be evaluated from the {\it classical}
dimer model, which has been extensively studied \cite{MFisher}. In
this section we address how those results fit into the framework
discussed here - for the case of the single layer bipartite
lattices. For example we may ask in the case of the honeycomb lattice
where a transition through a multicritical point can obtain by tuning
two parameters, whether the RK wavefunction corresponds to the the
ground state wave function of any of these fixed points. In fact, we
will conclude that while the RK points in both the square and the honeycomb
lattice cases can reproduce critical properties of some point on the
line of fixed points obtained after tuning a few parameters in the
generic models, they 
represent very special multicritical points in terms of their position
in the phase diagram of these generic models. Thus accessing these
points requires fine tuning an infinite number of independent
operators, which do not affect the critical properties but change the
phase structure in the immediate vicinity of the point (dangerously
irrelevant operators). This fine tuned nature of the RK
point can immediately be seen by noting that at the RK point, the
ground state wavefunction in each winding number (tilt) sector has
exactly the same energy. Reproducing this democratic treatment of all
winding number sectors within a height model representation of the RK
point will require tuning an infinite number of parameters to zero in
the bare Hamiltonian - even though they may be associated with operators
that are (dangerously) irrelevant at the critical point. 

\begin{figure}
\includegraphics[width=8.7cm]{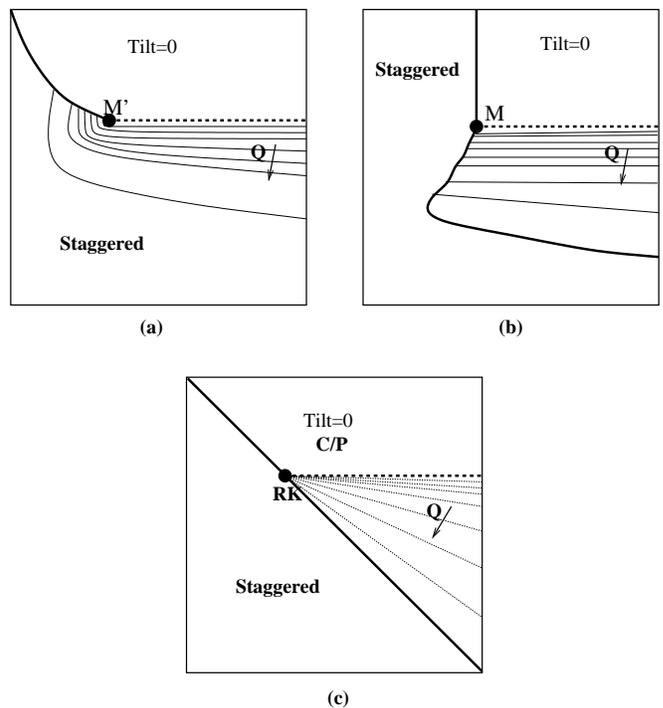}
\vspace{0.1in}
\caption{Schematic depiction of the phase diagram of VBSs, the
vertical axis in the plots is roughly the the parameter $\rho$ and the
staggered state has the maximum tilt. The generic phase diagrams
expected for the {\it bilayer} honeycomb lattice are shown in (a) and
(b) (and also of the single layer honeycomb lattice after tuning one
parameter, the cubic term, to zero). (a) The continuous transition
(shown with the dashed line) is from a zero tilt phase to a region
where the tilt (${\bf Q}$) exhibits a a devil's staircase structure,
contours of equal tilt (thin solid lines) are shown. The critical line
ends in a multicritical point $ M'$ beyond which the transition is
first order (solid line). The horizontal axis here may be thought of
as the coefficient of the quartic term. (b) An alternate scenario,
there is again a continuous transition to a region with the devil's
staircase structure for the tilt. Here however, there is a
multicritical point $M$ that is adjacent to the staggered state which
could control a zero tilt to staggered state transition. The
horizontal axis here may be thought of as the energy cost of the
maximally tilted (staggered) state. (c) The RK point for the single
layer square/honeycomb lattice - an infinite number of parameters need
to be tuned to access this plane. Exact degeneracy of the different
winding number sector ground states implies that states with arbitrary
tilt lie infinitesimally close to the RK point as shown.}
\label{RKpt}
\end{figure}

If the RK point requires fine tuning infinitely many parameters, one
may ask how it is accessed so readily in the quantum dimer model. The
reason is that the dimer model usually contains only single plaquette
terms, where the RK point is accessed by tuning just one
parameter. However, including processes that involve several
plaquettes will require tuning increasingly larger number of
parameters to obtain the equal dimer superposition of the RK
point. Generically the dimer representation of spin systems will
include processes that involve arbitrarily large number of
plaquettes. Although these decay with increasing size, they are
generally non vanishing, and infinite fine tuning will be then be
required to reach the RK point.

For the square lattice, the field theory that reproduces asymptotic
properties of the RK point was studied by Henley \cite{Henley}. It was
found there that a quadratic action (\ref{scritical}) with
$K=\pi^2/4$, $\sigma=0$ gives the long distance dimer correlations at
the RK point (amusingly this is the parameter value at which the two
monopole insertion operator is marginal). The fact that the ground
states in all winding number sectors are degenerate at the RK point
implies that it is possible to access states with arbitrary values of
the tilt my moving infinitesimally away from the RK point. As a
result, the topology of the phase diagram near the RK point is as
shown in figure (\ref{RKpt}c). An infinite number of parameters have
been set to zero to access this plane of the phase diagram. While the
continuous transition from the zero tilt state leads to the devil's
staircase of tilted states, a direct transition from a zero tilt state
to the staggered state can be made by crossing the RK point. Thus the
RK point terminates the critical line, and sits on a line of first
order transitions. The special nature of the RK point, even as a
multicritical point, can be seen by comparing it to the generic
multicritical points of the bilayer honeycomb lattice (which require
tuning of two parameters to reach) denoted $M'$, $M$ in figures
\ref{RKpt}a,b. (These can also be thought of as higher order
multicritical points in the single layer honeycomb model, which will
require fine tuning of an additional parameter to reach the plane
depicted in the diagrams.)In the first scenario depicted in figure
\ref{RKpt}a, the multicritical point $M'$ terminates both the first
order transition line and the critical line, but it does not allow for
a direct transition from the zero tilt state to the staggered
state. In the other scenario figure (\ref{RKpt}b) the generic
multicritical point $M$ does allow for a direct transition from a zero
tilt state to the staggered state. However, it is fundamentally
distinct from the RK point, as can be seend from the difference in
topology of the phase structure around the $M$ and RK points, which
may be characterized as follows. We ask what phases may be accessed
from these points by a small change of bare parameters. For the
generic multicritical point $M$, the staggered state, the zero tilt
state, and a state with infinitesimal tilt can be accessed. For the RK
point however, besides the staggered and zero tilt state, states with
arbitrary values of the tilt can also be accessed in this manner, as
shown in the figure. This follows from the exact degeneracy of ground
states in different winding number sectors at the RK point.

\section{Conclusions}

In this paper we have shown that at least in certain instances
there are direct second order transitions between distinct valence
bond solid phases. Deconfinement obtains at the critical point
though both phases are conventional and confined. More precisely
the critical theory may be viewed as a gapless $U(1)$ gauge theory
with {\em irrelevant} instantons. The spin carrying excitations in
either phase are gapped spin-$1$ magnons or their composites.
Right at the critical point however the spin gap does not close
but the magnons decay into (gapped) spin-$1/2$ spinons.

All of this structure is very similar to the other examples of
`deconfined' quantum criticality studied in Ref.
\cite{Senthiletal}. However there are some interesting differences
in the details. First the critical points discussed in the present
paper are controlled by a critical fixed line with continuously
varying exponents. Second (upto a marginally irrelevant non-linear
operator) all points on this fixed line have a simple free field
description. The emergence of a (topological) global $U(1)$
symmetry that seems generic to deconfined quantum criticality
obtains for the transitions in this paper as well. However the
free field description implies an enormous number of further
emergent symmetries which are specific to these transitions.
Third, the dynamic scaling exponent $z$ is $2$ in the present
example (compared to $z = 1$ at the Neel-VBS transition). Finally
as detailed in Section \ref{devil} there is rich and interesting
structure with an infinite number of transitions on one side of
the deconfined critical point.

One off-shoot of these results is a clarification of the place of the
solvable RK point of quantum dimer models in a more general context of
phase transitions in quantum magnets. We find that the RK point
corresponds to a special multicritical point. 

We also showed that  these interesting phase transitions are 
best realized in bilayer spin-$1/2$ honeycomb lattice quantum antiferromagnets. It would be interesting for numerical 
work to explore specific spin models on such bilayers where these transitions can be accessed. 

Several extensions of our results are possible. It should be possible
to examine the role of various perturbing fields at the critical point
as well as the effects of finite temperature. It should also be
readily possible to examine transitions between different VBS phases
in higher spin quantum magnets. We leave these for future work.

Since the original submission of an electronic preprint of this work,
Ref.~\onlinecite{huse} appeared which considered some of the same
questions especially those regarding the single layered systems
discussed here. They correctly pointed out the first order nature of
the transition on the single layer honeycomb lattice, and the
`incompleteness' of the devil's staircase - which were the points of
disagreement with the earlier version of this work. These points have
been corrected and briefly mentioned in the present work. However in
contrast to Ref. ~\onlinecite{huse} we have focused here on the case
of the bilayer honycomb quantum magnet where a generically continuous
transition between VBS states {\it is} realised, and have studied some
of the interesting properties of this generic quantum critical point.

\section{Acknowledgements}

This work was sparked by conversations with S. Sondhi and Y.B. Kim.
We are particularly grateful to the former for sharing his insights on
quantum dimer models and the physics of the RK point.  TS would like
to thank O. Motrunich for a very useful prior collaboration on related
field theories. We also thank E. Fradkin and S. Sachdev for useful
discussions, and D. Huse and C. Henley for pointing out some errors in
a previous version of the manuscript.  This research was generously
supported by the National Science Foundation under grants DMR-0308945
(T.S.), DMR-9985255 (L.B.). We would also like to acknowledges funding
from the NEC Corporation (T.S.), the Packard Foundation (L.B.), the
Alfred P.  Sloan Foundation (T.S., L.B.), a Pappalardo Fellowship
(A.V.) and an award from The Research Corporation (T.S.).

\appendix
\section{Critical gauge theory}
\label{pht}
In this Appendix we will explicitly display the form of the continuum
action of the non-compact $U(1)$ gauge theory that corresponds to the
critical sine Gordon theory.  A general continuum Hamiltonian for a
non-compact $U(1)$ gauge theory in $D = 2+1$ dimensions takes the form
\begin{equation}
\label{qedham}
H = \int d^2 x \,  \rho |{\bf E}|^2 + K |{\boldsymbol \nabla}
  \times {\bf E}|^2 + B^2 + \cdots
\end{equation}
together with the Gauss law constraint
\begin{equation}
{\boldsymbol \nabla}\cdot{\bf E} = 0.
\end{equation}
Here ${\bf E}$ refers to the electric field while $B =
{\bf\hat{z}}\cdot{\boldsymbol \nabla} \times {\bf A}
=\epsilon_{ij}\partial_i A_j$ is the magnetic field. We work in the
Coulomb gauge ${\boldsymbol \nabla}\cdot{\bf A} = 0$. As usual the
components of the electric field $E_i$ $(i = x,y)$ and the
(transverse) components $a_i$ of the vector potential are canonically
conjugate: \be [E_i({\bf x}), A_j({\bf x'})] = -iP_{ij}({\bf x - x'})
, \ee where $P_{ij}$ has Fourier components $\delta_{ij} - \frac{k_i
  k_j}{k^2}$ and projects out the transverse component.

The Gauss law constraint may be solved by writing 
\be E_i =\epsilon_{ij}\partial_j\chi, \ee 
where $\chi$ is a scalar field.
Assume that the commutator of $\chi$ and $A_j$ takes the form 
\be
[\chi({\bf x}), A_j({\bf x'})] = if_j({\bf x -  x'}) .
\ee 
The
correct commutator between ${\bf E}$ and ${\bf A}$ is reproduced if
we impose 
\be \epsilon_{ik}\partial_k f_j({\bf x -  x'}) =-
P_{ij}({\bf x -  x'}). 
\ee 
This then implies the commutator 
\be
[\chi({\bf x}), B({\bf x'})] = i\delta({\bf x -  x'}), \ee 
so that
the magnetic field $B$ is conjugate to $\chi$. The Hamiltonian may
now be rewritten 
\be 
\label{Ham}
H = \int d^2 x\,  \rho |{\boldsymbol \nabla} \chi|^2 + K
\left(\nabla^2 \chi\right)^2 + B^2 \ee As $B$ and $\chi$ are
canonically conjugate we reproduce the continuum free field action
of the sine Gordon theory. Remembering that the critical theory
has $\rho = 0$ we can immediately read off from Eqn. \ref{qedham}
the continuum gauge theory Hamiltonian that describes the critical
point. This is readily diagonalized explicitly to find a quadratic
dispersing photon.

Amusingly, at $\rho=0$, the Hamiltonian (\ref{Ham}) exhibits a kind of
self `duality', obtained by exchanging the roles of $B$ and $\nabla^2
\chi$. That is, if we introduce the field $\phi$ such that
$B=\frac{\nabla^2 \phi}{\sqrt{K}}$, and its conjugate field $\Pi_\phi
= -\sqrt{K}\nabla^2 \chi$, then one obtains the same critical
Hamiltonian (\ref{Ham} with $\rho=0$) but with $K \rightarrow 1/K$ and
 $(B,\, \chi)\rightarrow (\Pi_\phi,\, \phi)$.


\begin{thebibliography}{x}


\bibitem{Hertz} J.~A.~Hertz, Phys. Rev. B {\bf 14}, 1165 (1976).

\bibitem{Senthiletal}
T. Senthil, A. Vishwanath, L. Balents, S. Sachdev, and
M.~P.~A. Fisher, {\tt cond-mat/0311326}.

\bibitem{Senthiletal1}
T. Senthil,  L. Balents, S. Sachdev, A. Vishwanath, and
M.~P.~A. Fisher, {\tt cond-mat/0312617}.

\bibitem{MV} O.~Motrunich and A.~Vishwanath {\tt cond-mat/0311222}.

\bibitem{Henley} C.L.~Henley, J. Stat. Phys. {\bf 89}, 483
(1997). C.L.~Henley, {\tt cond-mat/ 0311345}.

\bibitem{MoSoFr}  R.~Moessner, S.~L.~Sondhi, and E.~Fradkin,
Phys. Rev. B {\bf 65}, 024504 (2002).

\bibitem{qdm}  D.~S.~Rokhsar and S.~A.~Kivelson, Phys. Rev. Lett. {\bf 61},
2376 (1988).

\bibitem{huse} E. Fradkin, D.A. Huse, R. Moessner, V. Oganesyan, and
S. Sondhi, {\tt cond-mat/0311353}.


\bibitem{Grinstein} G.~Grinstein, Phys. Rev. B {\bf 23}, 4615 (1981).

\bibitem{ReSaSuN} N.~Read and S.~Sachdev, Phys. Rev. Lett. {\bf 62}, 1694
  (1989);  N.~Read and S.~Sachdev, Phys. Rev. B {\bf 42}, 4568 (1990).

\bibitem{Frbook} E. Fradkin and S. A. Kivelson, Mod. Phys. Lett. B 4, 225 (1990); 
E. Fradkin, {\em Field theories of Condensed Matter Systems}, Perseus Books (1991).

\bibitem{Ardonne} E. Ardonne, P. Fendley and E. Fradkin, {\tt
cond-mat/ 0311466}.

\bibitem{Courtney} C. Lannert, M.~P.~A.~Fisher, and T.~Senthil,
Phys. Rev. B {\bf 63}, 134510 (2001).



\bibitem{Poly} A.~M.~Polyakov, {\em Gauge Fields and Strings}, Hardwood
Academic Publishers (1987).

\bibitem{subirrev}  For a recent review see S.~Sachdev, Annales Henri Poincare {\bf 4}, 554 (2003).


\bibitem{hald88} F.~D.~M.~Haldane, Phys. Rev. Lett. {\bf 61}, 1029 (1988).


\bibitem{sp} S.~Sachdev and K.~Park, Annals of Physics, N.Y. {\bf
    298}, 58 (2002).

\bibitem{note1} This is most readily seen in a $CP^1$ description in
which a skyrmion corresponds to $2\pi$ flux of the $U(1)$ gauge field
seen by the spinons.

\bibitem{Aubry} S.~Aubry and P.~Y.~Le Daeron, Physica D {\bf 8}, 381 (1983).

\bibitem{MFisher} M.~E.~Fisher and J.~Stephenson, Phys. Rev. 132, 1411
(1963).


\end{thebibliography}
\end{document}